\documentclass[12pt]{article}
\usepackage{arxiv}
\usepackage{comment}
\RequirePackage{amsthm,amsmath,amsfonts,amssymb}
\RequirePackage[authoryear]{natbib}
\usepackage{multirow}
\RequirePackage[colorlinks,citecolor=blue,urlcolor=blue]{hyperref}
\RequirePackage{graphicx}

\usepackage{siunitx}

\title{Climate Change in Austria: Precipitation and Dry spells over the last 60 years}

\author{Corinna Perchtold \\
  Johannes Kepler University\\
  Austria, 4040 Linz\\
  \texttt{corinna.perchtold@jku.at} 
}

\begin{document}

	\maketitle




\section{Abstract}
This study unveils localised changes in Austria's precipitation patterns, often missed by broader assessments, by comparing the 1961–1990 and 1991–2020 climate normal periods on a high resolution 2$\times$2 km grid. Our extended model explicitly accounts for diverse topographical influences, including slope, aspect, and a monthly-varying elevation effect, when analysing monthly normals of mean precipitation and maximum daily sums, as well as maximum dry spell lengths.
We found that while mean precipitation generally declined early in the year, it notably increased in March, September, and October (up to +50\%). In contrast, the maximum duration of dry spells extended significantly in January, February, and June, particularly in the southern regions (up to +30\%). Maximum daily precipitation amounts surged in late summer and autumn (up to +30\%). This research offers a transferable modelling approach for understanding critical shifts, vital for climate adaptation both within Austria and globally.

\section{Introduction}

Climate change analysis relies on defined 'climate normal periods' or 'standard reference periods', which are 30-year intervals established by the World Meteorological Organisation for reasons of international comparability and to serve as climatic baselines.\footnote{\url{https://community.wmo.int/en/activity-areas/climate-services/climate-products-and-initiatives/wmo-climatological-normals}}
 As of 2021, the new climate normal period is 1991–2020, succeeding the 1961–1990 period. This paper utilises this transition to conduct a detailed comparison of precipitation patterns between these two consecutive climate normal periods in Austria, addressing the main goal articulated in \cite{KlimaAut20}: 'To identify how the climate has evolved over the most recent three decades, pinpointing significant changes in key climate indices and exploring their implications for the environment and society in Austria.'

To achieve a more in-depth understanding of the evolution of climate, our study investigates local changes in three precipitation characteristics across Austria on a high resolution 2$\times$2 km grid and with publicly available data. This granular approach is particularly crucial given Austria's mountainous topography, where the Alps dominate the landscape. Mountains are recognised as climate hotspots where changes may precede or intensify those occurring in other areas, underscoring the great interest in precisely quantifying (local) precipitation shifts in such regions. 
Since past assessments, especially for small countries like Austria, often overlooked elevation-dependent changes in precipitation (\cite{Pepin2022}), and given that precipitation is a complex physical process with strong spatial and temporal dependencies (\cite{Fuentes2008}, \cite{Wang2015}), our effective model must account for these inherently region-specific factors.

Specifically, we examine three precipitation characteristics on a monthly timeline. 
Firstly, we consider monthly mean precipitation sums. A slight decrease in mean precipitation has been observed in Austria since the 1970s, see \cite{APCC14} or \cite{Hiebl2018}.
While resources like the Climate Change Knowledge Portal (providing precipitation climatology for Austrian states based on historical data produced by the Climatic Research Unit at the University of East Anglia for the periods 1961-1990 and 1991-2020) offer broad insights, our study aims for a more local investigation of these changes.\footnote{\url{https://climateknowledgeportal.worldbank.org/country/austria/climate-data-historical}}
The authors of \cite{Kotlarski2023} noted that the projected Alpine climate change simulated with the EURO-CORDEX initiative indicated a decrease in Alpine summer precipitation but an increase in winter, with summertime mean reduction driven by fewer but more intense wet days. 
We are therefore interested in examining the changes in Austria's mean precipitation normals over the last 60 years,
 and whether similar patterns emerge across all three of our precipitation indices.

Secondly, we model monthly  maximum daily precipitation sums.  
In \cite{Ombadi2023}, the authors discussed the amplification of maximum precipitation in high-altitude regions like the Alps, attributed to climate change. They highlighted a positive correlation between altitude and maximum precipitation, focusing particularly on the impact of warming-induced shifts from snow to rain. 
Their findings supported the hypothesis that precipitation extremes were shifting towards the colder seasons. 
For a more in-depth investigation of the rarity of extreme events, we compute 20-year return values representing precipitation thresholds expected to be surpassed, on average, once every two decades.
Further details on the statistical modelling of extreme values can be found in \cite{Coles2001}. 
The critical need for more practical research on extreme value theory and extreme climate phenomena in Austria is also underscored by the Central Institute of Meteorology and Geodynamics in Austria, \href{https://www.zamg.ac.at/cms/de/klima/informationsportal-klimawandel/standpunkt/klimazukunft/alpenraum/starkniederschlag}{GeoSphere}. They projected an increase in precipitation day intensity, with eastern Austria being particularly affected (\cite{Hiebl2018}).

Finally, we look at the monthly maximum length of a dry spell in Austria. Even if these are not a major problem in Austria at the moment, they still represent a future risk as they are likely to occur more frequently and with greater severity, according to \cite{Auer2005}. Various studies addressed this problem from a risk management perspective, as \cite{Leitner2020} or \cite{Palka2020}.
We consider the interruption of rainy days by a so-called dry spell, i.e., a sequence of consecutive dry days. The number of days defines the length of the dry spell and we choose the maximum length per month. 
The Austrian Federal Ministry of Agriculture, Forestry, Regions, and Water Management published the study 'Austria's Water Treasure' in 2023 (\cite{Ueberreiter23}), highlighting the rise in prolonged dry periods, such as those in 2015 and 2018, alongside increasing extreme rainfall events. They emphasised the need for further investigation to support policy preparedness.
 This aligned with findings from \cite{Heinrich2012}, who characterised the Greater Alpine Region as a challenging transition zone between wetter and dryer climates, underscoring the importance of in-depth, regionally-specific drought assessments for climate change understanding and water shortage preparedness.

Annual climate reports (\cite{KlimaAut19}, \cite{KlimaAut20}, \cite{KlimaAut21}, and  \cite{KlimaAut22}) compared drought indices across altitude zones (below 500m, 500-1000m, and up to 2000m) using the 1961-1990 reference period. They observed that the length of dry spells decreases with altitude. While these reports noted an increase in the total duration of dry spells compared to their reference period, they did not provide insights into specific spatial or monthly patterns, a gap our study aims to address.

In this paper, we aim to bridge existing knowledge gaps by providing a more thorough local analysis of three different precipitation indices, explicitly accounting for elevation dependence, and comparing normals across the last two standard reference periods on a monthly basis.
To achieve this goal, our approach is divided into the following tasks:
\begin{itemize}
\item Extend the model proposed in \cite{Cameletti2013} to explicitly capture non-stationary spatial and spatio-temporal effects, alongside enhanced topographical influences (including slope, aspect, and an interaction term between elevation and month) on Austrian precipitation patterns, utilising publicly available data.
\item Infer and present the changes in monthly normals of these precipitation characteristics, illustrating their evolution across the 1961-1990 and 1991-2020 climate normal periods.
\end{itemize}

The addition of modelling elevation and month as an interaction term is particularly important in our setting, as it allows the model to estimate a unique effect of altitude for each month, reflecting the variations in atmospheric conditions and precipitation throughout the year (\cite{Dura2024}, \cite{Benoit2024}). The slope and aspect are included to find out how precipitation changes with slope orientation 
and how this leads to differential precipitation patterns from north to south and east to west (\cite{Beullens2014}).

By focusing on monthly predictions of precipitation normals, our study addresses a key limitation of existing climate reports, which typically present only annual differences in precipitation indices. This coarse temporal resolution overlooks important seasonal variations that are critical for understanding climatic change.
Our resulting difference maps of the climate normal periods, showing monthly changes across Austria’s administrative districts, offer a more detailed spatial and temporal perspective than commonly found in established climate assessments, including the 2014 IPCC Assessment Report (\cite{APCC14}), publications by the Climate Change Center Austria (e.g., \cite{KlimaAut17, KlimaAut18, KlimaAut19, KlimaAut20, KlimaAut21, KlimaAut22}), and the Climate Change Knowledge Portal (\cite{Charney2018}).

Inference is done with INLA, (\cite{Rue2009}), in combination with the PARDISO solver.\footnote{\url{www.panua.ch}} This approach is readily accessible in the free software \texttt{R} through the \texttt{R-INLA} package.\footnote{\url{https://www.r-inla.org/download-install}}
Similar work on spatio-temporal variations in precipitation patterns with INLA is done, e.g., in \cite{Marques2020}, \cite{Vandeskog2022}, \cite{Mahmoudi2021}.

The paper is organised as follows. Section 2 presents the data. In Section 3, we introduce our model, present the distributional assumptions of the precipitation characteristics and explain the INLA approach. Section 4 presents the monthly relative changes in the posterior estimates between the two climate normal periods, the parameter estimates, and the cross-validation. In Section 5 we discuss the results and outline potential directions for future research and Section 6 concludes the paper.

\section{Material}
The Austrian Central Institute for Meteorology and Geodynamics, called GeoSphere, offers a data hub for the weather data that they have collected for any region in Austria and period in the past.\footnote{\url{https://data.hub.geosphere.at/}} This study is based on daily precipitation data (in mm) from all monitoring stations in Austria.

\subsection{Spatial domain}\label{sec:Austria}
Austria has a broadly diversified topology, ranging in longitude from 10\textdegree- 17\textdegree\, and latitude from 46\textdegree- 49\textdegree. Whereas the western and middle parts of Austria are rather dominated and shaped by the Alps, the area from north over east to south is predominantly hilly and flat. In fact, 2/3 of Austria is covered by mountains, with elevation above sea level varying between 112 and 3750 meters, see Figure \ref{fig:elev_map}.

\subsection{Precipitation data and monitoring sites}
\begin{figure}[h]
\centering		\includegraphics[scale=0.45]{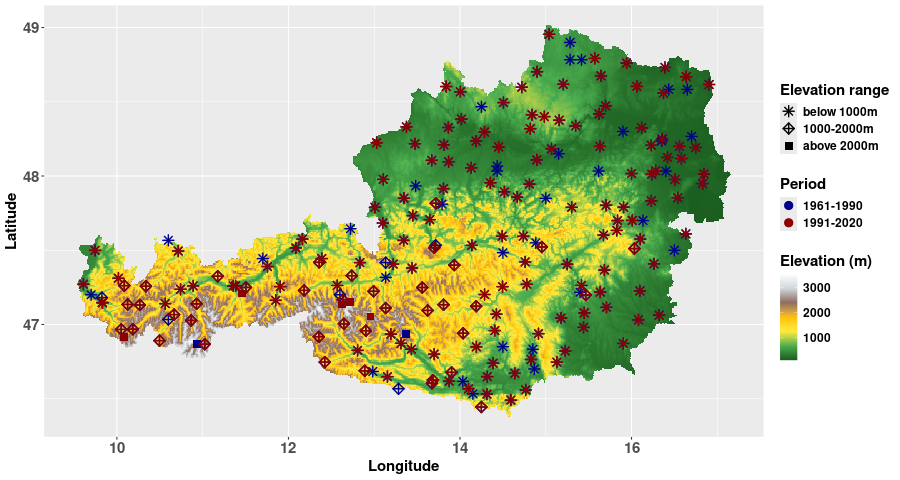}
		\caption{Elevation map of Austria with spatial distribution of monitoring stations.}
		\label{fig:elev_map}
\end{figure}

To examine how precipitation patterns changed over the last 60 years, we analyse data over two climate normal periods: 1961-1990 and 1991-2020.
Daily precipitation data for these periods are obtained from the GeoSphere monitoring network. Specifically, data from 224 stations active in the earlier period and 181 in the later one are utilised. 
Figure \ref{fig:elev_map} displays the spatial distribution of monitoring stations across Austria for both time periods. The stations are differentiated by elevation (below 1000m, 1000-2000m, and above 2000m) and are overlaid on an elevation map to provide geographical context. It also illustrates how GeoSphere relocated these stations over time.

For each monitoring station, we calculate three monthly precipitation characteristics from the daily data:
\begin{itemize}
    \item \textbf{Mean precipitation sum:} The average precipitation sum for each month.
    \item \textbf{Maximum daily precipitation sum:} The highest recorded daily precipitation sum for each month.
    \item \textbf{Maximum length of a dry spell:} The longest duration of consecutive dry days within each month.
\end{itemize}
A dry day is defined as a day with precipitation less than or equal to 0.1 mm.

The derived monthly values are then used to generate monthly "normals" (long-term means) for each period. These 30-year monthly normals provide a baseline for comparing long-term changes in precipitation characteristics between the two periods. Figure \ref{fig:normals} presents these monthly normals, highlighting the differences and similarities across the two periods for each characteristic.

In terms of data completeness, the dataset for the years 1961-1990 has approximately 15.8\% missing daily precipitation records, while the one from 1991-2020 has 17.3\% missing data.

\subsection{Covariates}
\begin{figure}[h!]
    \centering
    \includegraphics[scale=0.23]{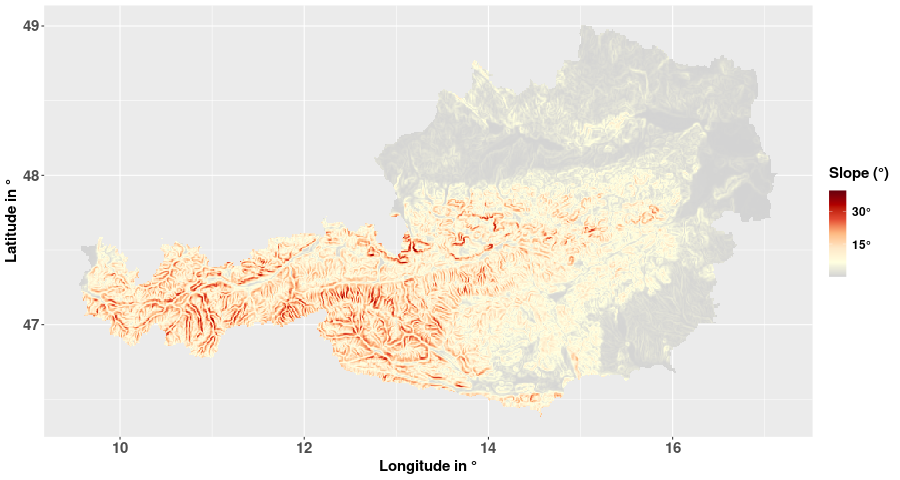}
        \includegraphics[scale=0.23]{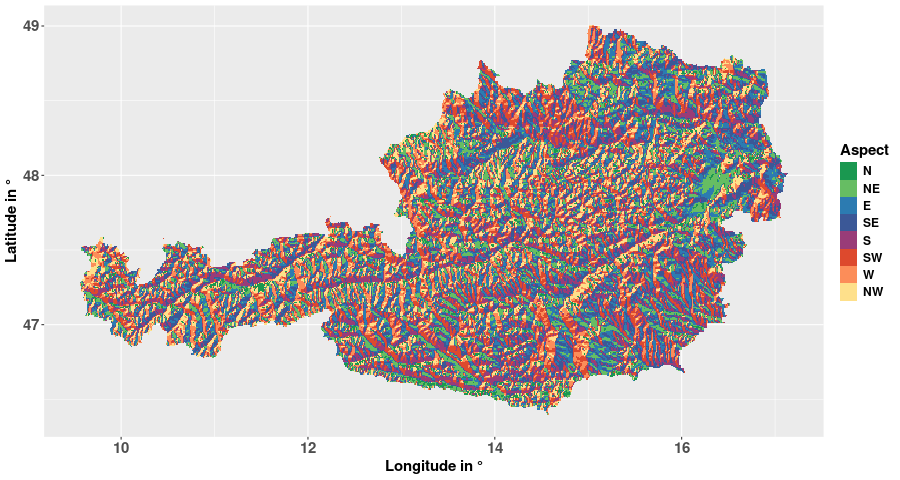}
    \caption{The topographic covariates slope (left) and aspect (right) of Austria.}
    \label{fig:topographic}
\end{figure}

Elevation is a fundamental topographical covariate accounting for orographic effects. To capture the dynamic and seasonally-dependent nature of its influence in the Austrian Alps, we include an interaction term between elevation and month (Elevation:Month). 
We scale elevation to kilometres for improved model interpretability. The elevation map is downloaded from the Global Administrative Areas (GADM) website.\footnote{\url{https://gadm.org/}}
After obtaining the GADM elevation map, we compute the aspect and slope from it with the \texttt{terra} package in \texttt{R}. The slope defines the steepness of the Austrian surface, by measuring the rate of elevation change across a horizontal distance. It is typically given as an angle, measured in degrees (\textdegree).
The aspect is another important topographic parameter since it refers to the direction in which a slope faces. It is measured in degrees (\textdegree), clockwise from 
north (0\textdegree) to east (90\textdegree) to 
south (180\textdegree) to 
west (270\textdegree) and back to north (360\textdegree).  Figure \ref{fig:topographic} shows the resulting topographic covariates. Both elevation, slope and aspect data are then integrated into our GeoSphere datasets.

\section{Method}
We propose a generalised linear mixed model, detailed in Subsection \ref{sec:gam}, which incorporates spatial and spatio-temporal random effects. This flexible model accommodates diverse types of precipitation observations (Subsection \ref{sec:data_dist}), enabling the estimation of monthly normals for various precipitation characteristics and a comprehensive comparison of patterns between the 1961-1990 and 1991-2020 reference periods. Model fitting is performed using the INLA framework (Subsection \ref{sec:INLA}), with hyperparameters specified in Subsection \ref{sec:priors}.

\subsection{Generalised linear mixed model}\label{sec:gam}
We model the monthly precipitation data $y(s_i, t_k)$ as a realisation of a spatio-temporal process defined over continuous space and time:
\[
\{y(s, t): s \in \mathbb{R}^2,\, t \in \mathbb{R}^+ \}.
\]
Measurements are taken at locations $s_i$, with either 224 or 181 active stations in the early or late 30-year period, respectively. Time is indexed monthly from $t_1=\text{January}$ to $t_{12}=\text{December}$. 

Conditioned on the mean $\mu(s_i, t_k)$, we assume that the data of each precipitation index follow a (non-normal) distribution $\pi(\cdot)$, i.e.,
\[
y(s_i, t_k) \mid \mu(s_i, t_k) \sim \pi(\mu(s_i, t_k)).
\]
The link function $g(\cdot)$ relates the mean to a structured additive predictor via 
\[
g(\mu(s_i, t_k)) = \eta(s_i, t_k).
\]
\noindent
The precipitation model is then defined in terms of the structured additive predictor $\eta(s_i, t_k)$, incorporating both fixed effects and random effects to account for unobserved sources of variability. For the fixed effects, we use the previously described topographic covariates.
The aspect and elevation are standardised, such that the coefficients can directly be interpreted as effect sizes.
Then the predictor is given by
\begin{equation}
\eta(s_i, t_k) = \alpha + \gamma_1\, \text{scale(Elevation):Month} + \gamma_2\, \text{scale(Aspect)} +\gamma_3\, \text{Slope} + \delta(s_i) + z(s_i, t_k),
\label{eq:gam}
\end{equation}
where:
\begin{itemize}
    \item $\alpha$ is the intercept,
    \item $\gamma_1$, $\gamma_2$ and $\gamma_3$ are regression coefficients,
    \item $\delta(s_i)$ is a spatial Gaussian random field (GRF) capturing time-invariant spatial effects,
    \item $z(s_i, t_k)$ is a spatio-temporal GRF capturing monthly variability.
\end{itemize}
\noindent
Temporal dependence in $z(s_i, t_k)$ is introduced via an AR(1) process over months $k = 2,\ldots,12$:
\begin{equation*}
z(s_i, t_k) = a z(s_i, t_{k-1}) + w(s_i), \quad \lvert a\rvert <1
\label{eq:ar1}
\end{equation*}
where $a$ denotes the temporal autocorrelation coefficient to be estimated, and $w(s_i)$ is a spatial innovation term.
We assume
\begin{itemize}
  \item   $\boldsymbol{ \delta} \sim \mathcal{N}(\boldsymbol{0}, \boldsymbol{\Sigma}_S)$, 
    \item $(z(s_1, t_1), \ldots, z(s_n, t_1))^\top \sim \mathcal{N}\left(\boldsymbol{0}, \frac{\boldsymbol{\Sigma}_S}{1 - a^2} \right)$ and $\boldsymbol{w} \sim \mathcal{N}(\boldsymbol{0}, \boldsymbol{\Sigma}_S)$, 
\end{itemize}
where $\boldsymbol{\Sigma}_S$ is the spatial covariance matrix based on a non-stationary isotropic Matérn covariance function, following \cite{Ingebrigtsen2014}.
To account for the effect of elevation across Austria, we model the marginal precision $\tau(s_i)$ and range $\kappa(s_i)$ parameters of the Matérn covariance as spatially varying log-linear functions:
\begin{align}
\log(\tau(s_i)) &= \theta_1^\tau + \text{Elevation}_i\cdot\theta_2^\tau, \label{eq:space-varying-parameter}\\
\log(\kappa(s_i)) &= \theta_1^\kappa + \text{Elevation}_i\cdot\theta_2^\kappa. \label{eq:space-varying-parameter2} 
\end{align}
When the marginal precision $\tau(s_i)$ varies spatially, it means that the uncertainty of the precipitation process changes with location. This is particularly relevant in mountainous countries like Austria, where, for example, precipitation might be inherently more variable in mountains than in valleys. Similarly, a spatially varying range $\kappa(s_i)$  indicates that the distance over which locations remain correlated is not constant across the study area, enabling the model to capture, for instance, a faster decay of correlation in steep terrain compared to flatter regions.
The parameters to be estimated are:
\[
\boldsymbol{\theta}^\delta = ( \theta_1^\tau, \theta_2^\tau, \theta_1^\kappa, \theta_2^\kappa), \quad \boldsymbol{\theta}^w =( \theta_1^\tau, \theta_2^\tau, \theta_1^\kappa, \theta_2^\kappa).
\]
This formulation allows for non-stationary behaviour in both the spatial field $\delta(s_i)$ and the innovations $w(s_i)$, capturing elevation-induced variability.

In contrast to the model in \cite{Cameletti2013}, we introduce an additional spatial term $\delta(s_i)$ in \eqref{eq:gam} to disentangle persistent spatial structure from monthly spatio-temporal dynamics. This separation enhances identifiability and improves estimation of each component in the model.

Both spatial fields $\delta(s_i)$ and $z(s_i, t_k)$ are represented as Gaussian Markov random fields (GMRFs) using the Stochastic Partial Differential Equation (SPDE) approach of \cite{Lindgren2011} and  \cite{Simpson2012}. This finite element-based method provides a computationally efficient representation of continuous spatial and spatio-temporal processes, while preserving the necessary conditional independence structure. Further details can be found in the Supplementary material, Section \ref{sec:Appendix}.

\subsection{Data models}\label{sec:data_dist}
  
Next, we specify the distributional assumptions for our three precipitation characteristics. For model fitting, we then use the likelihood functions provided by \texttt{R-INLA}, employing the default link functions associated with each distribution.  

		
\subsubsection{Model for monthly mean precipitation sums}\label{sec:avg}
Given that precipitation amounts are non-negative, the gamma distribution is selected for modelling monthly mean precipitation sums, which is a common choice according to the literature (\cite{MartinezVillalobos2019}, \cite{Zelalem2023}).


This distribution is typically parametrised by a shape parameter $\alpha$ and a rate parameter $\lambda$ (which is the inverse of the scale parameter), such that the conditional distribution of the data given $\alpha$ and $\lambda$ is $y(s_i,t_k)\mid \alpha, \lambda \sim \Gamma( \alpha, \lambda)$. The mean of this distribution is $\mu(s_i,t_k) = \frac{\alpha}{\lambda}$.
In the INLA framework, we explicitly connect these parameters to a precision parameter $\phi$. Specifically, we define the shape parameter as $\alpha = s\phi$, where $s>0$ is a fixed scaling factor. Consequently, the rate parameter is given by $\lambda = \frac{s\phi}{\mu(s_i,t_k)}$. 
The conditional distribution of $y(s_i,t_k)$ given its mean $\mu(s_i,t_k)$ and the precision parameter $\phi$ is then specified by:
$$y(s_i,t_k) \mid \mu(s_i,t_k), \phi \sim \Gamma( s\phi,  s\phi/\mu(s_i,t_k)),$$
where the mean $\mu(s_i,t_k)$ is  linked to the structured additive linear predictor $\eta(s_i,t_k)$ through a logarithmic link function:
$$\log(\mu(s_i,t_k)) = \eta(s_i,t_k).$$

\subsubsection{Model for monthly maximum daily precipitation sums}\label{sec:max}
The selection of the maximum daily precipitation sum for each station within each month, follows the block maxima approach, where each month resembles a block. 
 According to extreme value theory, for sufficiently large blocks, the distribution of these block maxima is approximately described by the generalised extreme value (GEV) distribution (\cite{Coles2001}). 
 However, due to its parameter-dependent support and limitations within our inference environment, we instead employ the blended generalised extreme value  (bGEV) distribution as proposed and reparametrised by \cite{CastroCamilo2022}. A comparison between the GEV and bGEV distribution is made in \cite{Krakauer2024}. 
 
 This reparametrisation
transforms the location, scale, and shape parameters from the GEV distribution, i.e., $(\mu,\sigma, \xi)$ to the location, spread, and tail parameter of the bGEV distribution $(q_\alpha, s_\beta, \xi)$.
Here, the location parameter $q_\alpha$ represents the $\alpha$-quantile, with $0<\alpha<1$, the spread parameter $s_\beta$ is defined as the quantile range, i.e., $s_\beta=q_{1-\beta/2}-q_{\beta/2}\in\mathbb{R}^+$, and the tail parameter is restricted to $\xi\in[0, \infty)$.  For our specific model, we allow $s_\beta$ to vary with elevation and initialise $\xi$ with $\xi=0.1$. Based on \cite{Vandeskog2022}, we set $\alpha=0.5$ and $\beta=0.8$. Consequently, the spread parameter $s_{0.8}=q_{0.6}-q_{0.4}$ represents the range of the central 20\% of the distribution and the effects of the covariates are described over the median of the data distribution, i.e., $q_{0.5}$. The conditional distribution for the monthly maximum daily precipitation sums $y(s_i,t_k)$ is given by
\[
y(s_i,t_k)\mid q_{0.5}, s_{0.8},\xi\sim \mbox{bGEV}( q_{0.5}, s_{0.8}, \xi).
\]
The median $q_{0.5}$ is linked to the structured additive linear predictor $\eta(s_i,t_k)$ through the identity function, i.e., $$\operatorname{id}(q_{0.5})=\eta(s_i,t_k).$$ 
    
\subsubsection{Model for the maximum length of a dry spell per month}\label{sec:dry}

In the literature the length of dry spells is often modelled through the negative binomial distribution, see the case study for Greece (\cite{Anagnostopoulou2003}) or Tunisia (\cite{Mathlouthi2012}). 

This distribution is typically parametrised by the dispersion parameter $n>0$ and probability $p\in [0,1]$, such that $y(s_i,t_k)\mid n,p \sim \text{NB}(n, p)$. The mean of this distribution is $\mu(s_i,t_k) = n\frac{(1-p)}{p}$.
The probability $p$ can be expressed by  $p = \frac{n}{\mu(s_i,t_k)+n}$.

The conditional distribution of $y(s_i,t_k)$, given its mean $\mu(s_i,t_k)$ and the dispersion parameter $n$, is then specified by:
$$y(s_i,t_k) \mid n, \mu(s_i,t_k) \sim \text{NB}\left (n, \frac{n}{\mu(s_i,t_k)+n} \right),$$
where the mean $\mu(s_i,t_k)$ is linked to the structured additive linear predictor $\eta(s_i,t_k)$ through a logarithmic link function:
$$\log(\mu(s_i,t_k)) = \eta(s_i,t_k).$$
		
\subsection{Estimation with INLA}\label{sec:INLA}

Data analysis and modelling are performed using \texttt{R} (\cite{RTeam2011}) and the \texttt{R-INLA} package developed in \cite{Lindgren2015}.\footnote{\url{https://www.r-inla.org/}}
Our goal is to infer the monthly normals of each time period 
on a high resolution map with a grid size of 2$\times$2 km.  All computations are performed on a dual Intel Xeon CPU E5-2687W with $750$GB RAM.

This inference procedure considers the precipitation values at prediction points missing. Consequently, missing data from monitoring stations are also estimated, and therefore not problematic. 
The hyperparameters for each precipitation characteristics are:
		\begin{itemize}
			\item for monthly mean precipitation sums $\boldsymbol{\psi}=\{ \phi, \boldsymbol{\theta^\delta}, \boldsymbol{\theta^w},a \}$,
			\vspace{2mm}
			\item for monthly maximum daily precipitation sums $\boldsymbol{\psi}=\{  s_\beta, \xi, \beta_1,\boldsymbol{\theta^\delta}, \boldsymbol{\theta^w}, a \}$,
			\vspace{2mm}
			\item for the monthly maximum length of a dry spell $\boldsymbol{\psi}=\{n, \boldsymbol{\theta^\delta}, \boldsymbol{\theta^w}, a \}$,
		\end{itemize}
with $\boldsymbol{\theta^\delta}$ and $ \boldsymbol{\theta^w}$  the vector of weight parameters of the spatially varying parameters $\kappa(s_i)$ and $\tau(s_i)$, explained in Equation (\ref{eq:space-varying-parameter}) and (\ref{eq:space-varying-parameter2}), $\beta_1$ the regression coefficient of the spread $s_\beta$ and the residual parameters coming from Subsection \ref{sec:data_dist}. 

The objectives of the Bayesian computations are the marginal posterior distributions for each of the elements of the latent vector $\boldsymbol{\zeta}=(\alpha, \gamma_1, \gamma_2, \gamma_3,\boldsymbol{\delta}, \boldsymbol{z})$ with $\boldsymbol{\delta}=\{\delta(s_i)\}$ and $\boldsymbol{z}=~\{z(s_i,t_k)\}$ and the hyperparameter vector $\boldsymbol{\psi}$ conditioned on the data $\boldsymbol{y}=\{y(s_i,t_k)\}$:
\[
\begin{gathered}
\pi\left(\psi_\ell \mid \boldsymbol{y}\right)=\int \pi(\boldsymbol{\psi} \mid \boldsymbol{y}) \mathrm{~d} \boldsymbol{\psi}_{-\ell}, \\
\pi\left(\zeta_j \mid \boldsymbol{y}\right)=\int \pi\left(\zeta_j \mid \boldsymbol{\psi}, \boldsymbol{y}\right) \pi(\boldsymbol{\psi} \mid \boldsymbol{y}) \mathrm{d} \boldsymbol{\psi},
\end{gathered}
\]
where $j$ and $\ell$ are the length of the vector $\boldsymbol{\zeta}$ and $\boldsymbol{\psi}$, respectively. 
The INLA approach exploits the assumptions of the model to produce a numerical approximation to the posteriors of interest, based on the Laplace approximation (\cite{Tierney86}).
For more information on the approach, read \cite{Bakka2018}, \cite{Blangiardo2015}, \cite{Martins2013} or \cite{Rue2009}.

\subsection{Prior assumptions}\label{sec:priors}
To complete the model specification, we assign prior distributions to the elements of the latent field $\boldsymbol{\zeta}$ and hyperparameters $\boldsymbol{\psi}$. We adopt INLA's default priors for the intercept $\alpha$, the coefficients $\gamma_1$, $\gamma_2$ and $\gamma_3$, the precision parameter $\phi$, the dispersion parameter $n$, the spread and tail parameter $s_\beta$ and $\xi$ and the coefficient of the spread $\beta_1$.
The non-stationary Matérn parameters $ \theta_1^\tau, \theta_2^\tau, \theta_1^\kappa, \theta_2^\kappa$, introduced in (\ref{eq:space-varying-parameter}) and~(\ref{eq:space-varying-parameter2}), are assigned standard normal priors with zero mean and unit precision based on \cite{Ingebrigtsen2014} and \cite{Marques2020}. We assign a Log-Gamma prior  with shape parameter equal to 1 and rate parameter equal to 0.01 to the dispersion parameter $n$.
For the mean precipitation and maximum length of a dry spell models, we further specify a PC prior for the temporal AR($1$) correlation.
To express prior belief in strong temporal correlation, 
we set a $90\%$ probability of having a temporal correlation above $90\%$. For the maximum precipitation model, this expectation is adjusted to an $80\%$ probability of the correlation exceeding $70\%$ ( information on the PC prior can be found in   \texttt{inla.doc("pc.cor1")}).\footnote{Further documentation: \texttt{inla.doc("gamma")}, \texttt{inla.doc("bgev")}, \texttt{inla.doc("nbinom")}, \texttt{inla.doc("pc.gevtail")}.}


\section{Results}

This chapter presents a comprehensive visualisation of relative monthly differences in posterior normals from mean precipitation sums, maximum daily precipitation sums, 20-year return values of maximum daily precipitation sums, and the maximum length of dry spell across Austria, comparing the 1991-2020 to the 1961-1990 climatic reference periods.
The 20-year return values, estimated using \cite{Vandeskog2022} and the \texttt{R} package \texttt{inlaBGEV}, provide a robust measure of extreme daily precipitation, representing events expected once every two decades. This makes them well-suited for comparison across 30-year standard references periods, where such events occur on average 1.5 times, ensuring sufficient sampling of extremes for statistical analysis (\cite{Coles2001}).
Additionally, we present the relevant parameter estimates in Subsection \ref{sec:estimates} and cross-validation measures in \ref{sec:cv}. 

\subsection{Posterior estimates of precipitation characteristics}

\begin{figure}[h!]
\centering
    \includegraphics[scale=0.5, trim= 0 25mm 0 10mm,clip]{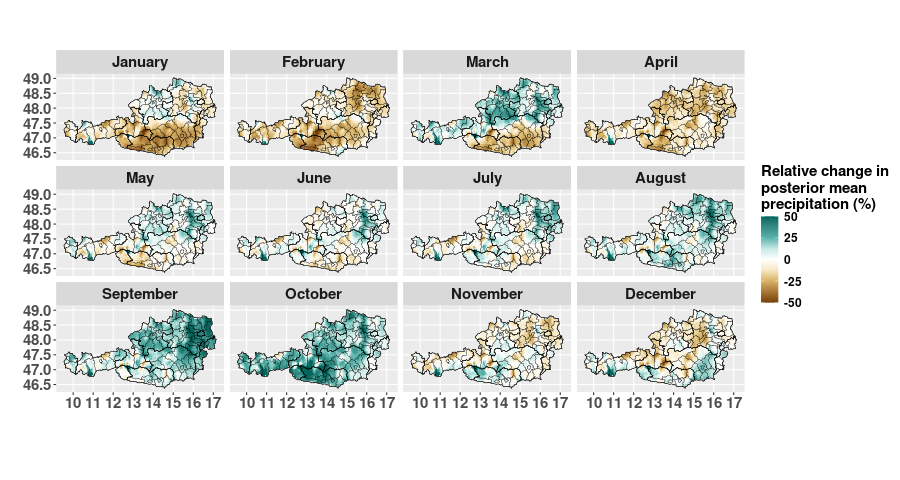}
    \caption{ Monthly relative change in posterior mean precipitation sum across Austria's administrative districts. The differences are calculated from the monthly normals of 1991-2020 and 1961-1990.}
    \label{fig:monthly_mean}
    \includegraphics[scale=0.5, trim= 0 25mm 0 10mm,clip]{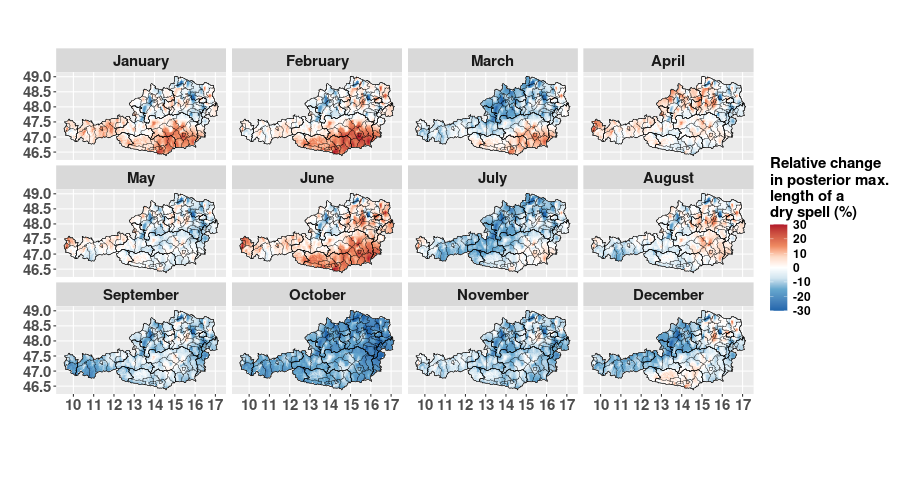}
    \caption{ Monthly relative change in posterior maximum length of a dry spell across Austria's administrative districts. The differences are calculated from the monthly normals of 1991-2020 and 1961-1990.}
    \label{fig:monthly_dry_spell}
        \end{figure}

\begin{figure}[h!]
\centering
    \includegraphics[scale=0.5, trim= 0 25mm 0 10mm,clip]{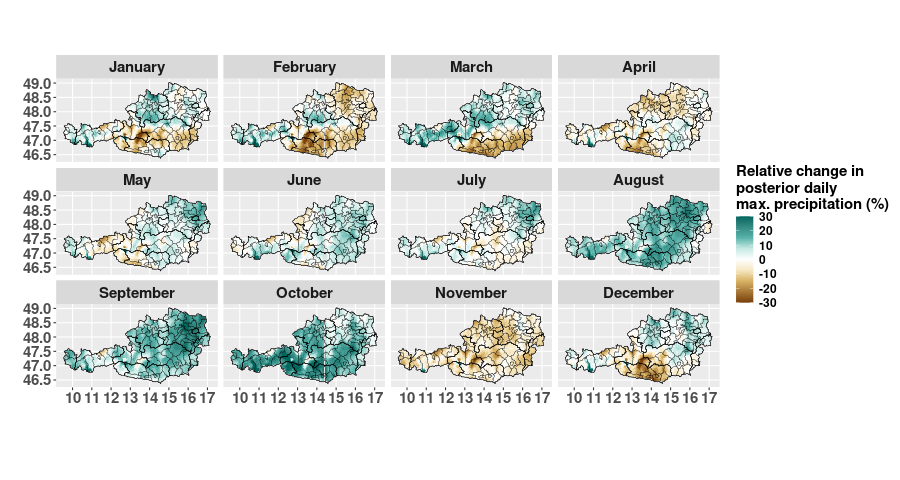}
    \caption{ Monthly relative change in posterior maximum daily precipitation sums across Austria's administrative districts. The differences are calculated from the monthly normals of 1991-2020 and 1961-1990.}
        \label{fig:monthly_daily_max}   \includegraphics[scale=0.5, trim= 0 25mm 0 10mm,clip]
        {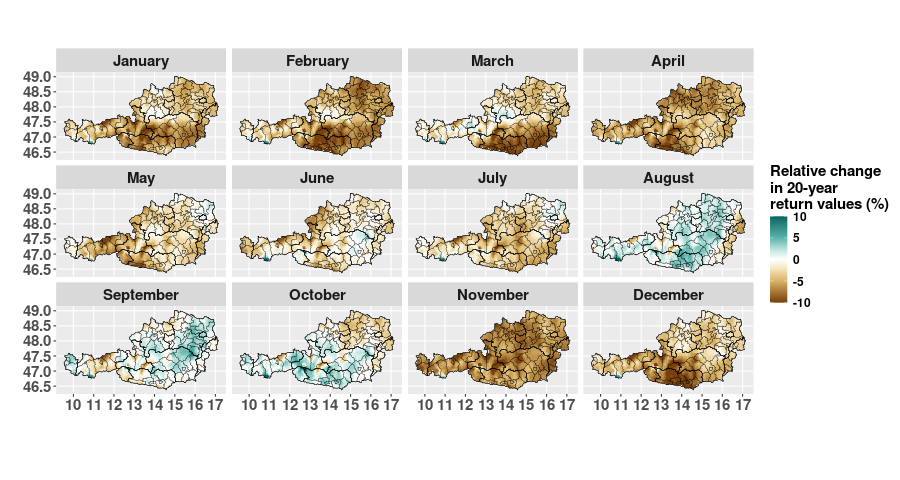}
    \caption{Monthly relative change in posterior 20-year return values across Austria's administrative districts. The differences are calculated from the monthly normals of 1991-2020 and 1961-1990.}
    \label{fig:monthly_max}
    \end{figure}

In the winter months of January and February, a consistent pattern of generally drier conditions is observed in the southern part of Austria. 
This trend, evidenced by Figures \ref{fig:monthly_mean}, \ref{fig:monthly_dry_spell} and \ref{fig:monthly_daily_max} 
includes reduced mean precipitation of up to $50\%$, increased maximum duration of dry spells of up to $30\%$, and a decrease in the magnitude of extreme daily precipitation events of up to $30\%$ in the period 1991-2020 compared to 1961-1990.
However, on the northern side of the Alps the monthly precipitation mean remains largely stable and the length of dry spells shows only minor changes but the intensity of the single extreme daily rainfall event within the months has, on average, increased. This suggests a pattern where the overall monthly precipitation amount is maintained, but it is increasingly delivered through more concentrated daily events.

Spring, from March to June, presents a regionally differentiated picture.  In March, mean precipitation shows mixed trends in the north ($\approx +25-50\%$ ) and south of the Alps ($\approx -25\%$), with a corresponding response in dry spell length (north $-20\%$, south $+20\%$). Extreme daily precipitation decreases in the south but the map indicates higher daily maximums in the northern parts and along the Alps.
In April, a nationwide trend towards reduced mean precipitation is observed, which in turn favours corresponding increases in dry spell length, primarily in the central to northern regions. This is accompanied by a reduction in the normals of daily maximum precipitation sums.
The pattern of trends observed for May suggests a concentration of mean precipitation increases alongside higher maximum daily precipitation sums, accompanied by a shortening of dry spells in northern Austria. 
A more large-scale trend is observed in June: the relative differences show a tendency towards longer dry spell normals and higher maximum daily precipitation normals in the southern and eastern parts (both up to $30\%$) from 1961-1990 to 1991-2020.

During the late spring and summer months of July and August, the mean precipitation maps indicate localised increases in the north-eastern part of Austria.
In turn, July displays a strong reduction in dry spell lengths of up to $30\%$ in most of the country.
August shows a particularly marked increase (up to $30\%$) in maximum daily precipitation sums across large parts of Austria, indicating that the most intense daily rainfall events in this month have, on average, become more severe in 1991-2020 compared to 1961-1990.
A more in-depth picture yield the 20-year return values in Figure \ref{fig:monthly_max}.
While a reduction in the intensity of such a 20-year return value is observed until July, for the period 1991-2020 compared to 1961-1990, the month of August indicates that the magnitude of such a rare daily event is increasing by up to $10\%$ for most of Austria. 
This suggests that while mean summer precipitation may be increasing in some areas, the risk of very intense, short-duration extreme daily precipitation events is rising in late summer for Austria.

The months of autumn, September and October, generally showcases a pattern of overall wetter periods across central, eastern, and southern Austria with a heightened intensity of both daily sums and 20-year return value events.
 This collectively suggests that autumn is becoming characterised by much higher precipitation volumes and notably shorter dry periods.
 
 While at the end of the year the mean precipitation largely remains unchanged with localised decreases, it still shows a reduction in dry spell length and maximum daily precipitation events from western to northern Austria. Only in December the northern and eastern part show a relative intensification in maximum daily precipitation sums in the later climate period.

In conclusion, these findings reveal a regionally differentiated picture in Austria's seasonal precipitation patterns. The country is experiencing drier winters alongside an increase in dry spell length and less intense daily extremes in the South. March and autumn months tend to become wetter with decreased drought risk. 
Conversely, late spring and summer show a significant increase in the intensity of extreme daily precipitation events especially in eastern and southern Austria. This highlights a climate becoming more extreme in its seasonality and intensity, clearly visible when examining the 20-year return values of maximum daily precipitation.

\subsection{Posterior estimates of model parameters}\label{sec:estimates}

To assess the statistical significance of the model parameters, we analyse their posterior means and the bounds of their 95\% credible intervals (CIs), specifically determined by the 2.5th and 97.5th percentiles of their posterior distributions. A parameter is not statistically significant if its 95\% CI includes zero.


\begin{table}[ht]
    \centering
    \label{tab:posterior_mean}
    \begin{tabular}{lccc}
        \hline
        \textbf{Variable} & \textbf{1961-1990: Mean} (\textbf{95\% CI}) & \textbf{1991-2020: Mean} (\textbf{95\% CI})\\
        \hline 
        \multicolumn{3}{l}{\emph{Fixed Effects}} \\
          $\quad$ Slope & 0.002\,\,\, (-0.006, 0.010) & -0.005\,\,\, (-0.015, 0.005)  \\
          $\quad$ scale(Aspect) & \textbf{-0.012}\,\,\, (-0.025, -0.001)  & \textbf{-0.026}\,\,\, (-0.042, -0.010)\\
        $\quad$ scale(Elevation): January &  \textbf{0.107}\,\,\, (0.077, 0.138) &  \textbf{0.136}\,\,\, (0.099, 0.174)\\
        $\quad$ scale(Elevation): February & \textbf{0.119}\,\,\, (0.089, 0.150)  & \textbf{0.161}\,\,\, (0.124, 0.198)  \\
        $\quad$ scale(Elevation): March & \textbf{0.135}\,\,\, (0.106, 0.166)  & \textbf{0.124}\,\,\, (0.088, 0.161) \\
        $\quad$ scale(Elevation): April & \textbf{0.101}\,\,\, (0.072, 0.132)  &  \textbf{0.107}\,\,\, (0.070, 0.143)\\
        $\quad$ scale(Elevation): May & \textbf{0.056}\,\,\, (0.026, 0.087)  &  \textbf{0.042}\,\,\, (0.005, 0.078)\\
        $\quad$ scale(Elevation): June & \textbf{0.036}\,\,\, (0.006, 0.066)  &  0.027\,\,\, (-0.010, 0.064)\\
        $\quad$  scale(Elevation): July  & 0.027\,\,\, (-0.003, 0.057)  & 0.025\,\,\, (-0.012, 0.062) \\
         $\quad$ scale(Elevation): August & \textbf{0.031}\,\,\, (0.001, 0.061) &  0.017\,\,\, (-0.020, 0.053)\\
         $\quad$ scale(Elevation): September  & \textbf{0.032}\,\,\, (0.003, 0.063)   &  0.022\,\,\, (-0.015, 0.058)\\
         $\quad$ scale(Elevation): October & 0.025\,\,\, (-0.004, 0.057)   &  0.012\,\,\, (-0.025, 0.048)\\
         $\quad$ scale(Elevation): November & \textbf{0.031}\,\,\, (0.001, 0.062)  &  \textbf{0.042}\,\,\, (0.004, 0.078)\\
         $\quad$ scale(Elevation): December & \textbf{0.090}\,\,\, (0.060, 0.122) &  \textbf{0.081}\,\,\, (0.043, 0.118)\\[.1cm]
     \multicolumn{3}{l}{\emph{Random Effects: Distribution }} \\
$\quad$ Precision-parameter $\phi$ & \textbf{3.014}\,\,\, (2.984, 3.045) & \textbf{2.852}\,\,\,  (2.820, 2.885) \\[.1cm]
        \multicolumn{3}{l}{\emph{Random Effects: Spatial field}} \\
          $\quad$  $\theta_1^\tau$  & 0.520\,\,\,  (-1.820, 3.285)  & -0.559 \,\,\, (-1.922, 0.475)  \\
         $\quad$ $\theta_2^\tau$  & \textbf{-0.556}\,\,\, (-0.982, -0.176)   & -0.065\,\,\,  (-0.503, 0.474) \\
         $\quad$  $\theta_1^\kappa$  & -0.075\,\,\, (-2.852, 2.270)  & \textbf{1.315}\,\,\,  (0.671, 2.149) \\
        $\quad$  $\theta_2^\kappa$ & \textbf{-0.686}\,\,\, (-1.066, -0.252)  & \textbf{-1.209}\,\,\,  (-1.641, -0.826) \\ [.1cm]
        \multicolumn{3}{l}{\emph{Random Effects: Spatio-temporal field}} \\
          $\quad$  $\theta_1^\tau$  & 0.116\,\,\, (-0.159, 0.390)  &  0.107\,\,\,  (-0.168, 0.411)\\
         $\quad$ $\theta_2^\tau$  & 0.043\,\,\, (-0.224, 0.326) & \textbf{-0.245}\,\,\,  (-0.489, -0.053) \\
         $\quad$  $\theta_1^\kappa$  & -0.127\,\,\, (-0.379, 0.127) & -0.070\,\,\,  (-0.325, 0.165) \\
        $\quad$  $\theta_2^\kappa$ & \textbf{-0.127}\,\,\, (-0.222, -0.031) & \textbf{-0.113}\,\,\,  (-0.211, -0.014)\\
        $\quad$ Temporal correlation $a$ & \textbf{0.831}\,\,\, (0.721, 0.903) & \textbf{0.873}\,\,\,  (0.819, 0.929)\\ [.1cm]
        \hline
    \end{tabular}
     \caption{Posterior estimates of fixed and random effects for mean precipitation sum normals. Statistically significant effects are highlighted in bold.}
\end{table}

\begin{table}[h!]
    \centering
    \label{tab:posterior_max}
    \begin{tabular}{lccc}
        \hline
        \textbf{Variable} & \textbf{1961-1990: Mean} (\textbf{95\% CI}) & \textbf{1991-2020: Mean} (\textbf{95\% CI})\\
        \hline 
        \multicolumn{3}{l}{\emph{Fixed Effects}} \\
          $\quad$ Slope & 0.053\,\,\, (-0.025, 0.131) & -0.030\,\,\, (-0.133, 0.074)  \\
          $\quad$ scale(Aspect) & \textbf{-0.156}\,\,\, (-0.286, -0.021)  & \textbf{-0.269}\,\,\, (-0.432, -0.106)\\
        $\quad$ scale(Elevation): January &  \textbf{1.087}\,\,\, (0.649, 1.532) &  \textbf{1.281}\,\,\, (0.782, 1.777)\\
        $\quad$ scale(Elevation): February & \textbf{1.178}\,\,\, (0.759, 1.607)  & \textbf{1.413}\,\,\, (0.919, 1.903)  \\
        $\quad$ scale(Elevation): March & \textbf{1.221}\,\,\, (0.800, 1.647)  & \textbf{1.270}\,\,\, (0.783, 1.756) \\
        $\quad$ scale(Elevation): April & \textbf{1.160}\,\,\, (0.731, 1.594)  &  \textbf{1.240}\,\,\, (0.755, 1.724)\\
        $\quad$ scale(Elevation): May & \textbf{1.180}\,\,\, (0.774, 1.591)  &  \textbf{0.937}\,\,\, (0.457, 1.415)\\
        $\quad$ scale(Elevation): June & \textbf{1.491}\,\,\, (1.080, 1.908)  &  \textbf{1.161}\,\,\, (0.674, 1.642)\\
        $\quad$  scale(Elevation): July  & \textbf{1.448}\,\,\, (1.040, 1.865)  & \textbf{1.256}\,\,\, (0.768, 1.737) \\
         $\quad$ scale(Elevation): August & \textbf{1.533}\,\,\, (1.119, 1.956) &  \textbf{1.320}\,\,\, (0.822, 1.809)\\
         $\quad$ scale(Elevation): September  & \textbf{1.106}\,\,\, (0.691, 1.526)   &  \textbf{1.147}\,\,\, (0.660, 1.629)\\
         $\quad$ scale(Elevation): October & \textbf{0.997}\,\,\, (0.559, 1.446)   &  \textbf{1.431}\,\,\, (0.949, 1.911)\\
         $\quad$ scale(Elevation): November & \textbf{0.966}\,\,\, (0.539, 1.398)  &  \textbf{0.942}\,\,\, (0.443, 1.440)\\
         $\quad$ scale(Elevation): December & \textbf{0.772}\,\,\, (0.335, 1.214) &  \textbf{1.126}\,\,\, (0.621, 1.629)\\[.1cm]
     \multicolumn{3}{l}{\emph{Random Effects: Distribution }} \\
$\quad$ Spread-parameter $s_\beta$ & \textbf{4.675}\,\,\, (4.621, 4.734) & \textbf{4.940}\,\,\,  (4.873, 5.001) \\
$\quad$ Tail-parameter $\xi$ & \textbf{0.146}\,\,\, (0.140, 0.153) & \textbf{0.107}\,\,\,  (0.100, 0.114) \\
$\quad$ Coefficient of spread $\beta_1$  & \textbf{0.150}\,\,\, (0.136, 0.164) & \textbf{0.143}\,\,\,  (0.130, 0.159) \\ [.1cm]
\multicolumn{3}{l}{\emph{Random Effects: Spatial field}} \\
 $\quad$  $\theta_1^\tau$  & -0.475\,\,\,  (-1.292, 0.268)  & -0.130 \,\,\, (-1.259, 1.338)  \\
 $\quad$ $\theta_2^\tau$  & -0.815\,\,\, (-2.044, 0.548)   & \textbf{-2.202}\,\,\,  (-3.259, -1.201) \\
 $\quad$  $\theta_1^\kappa$  & 1.004\,\,\, (-0.217, 2.495)  & \textbf{1.651}\,\,\,  (0.532, 3.121) \\
 $\quad$  $\theta_2^\kappa$ & \textbf{-1.624}\,\,\, (-2.977, -0.538)  & \textbf{-1.348}\,\,\,  (-1.973, -0.739) \\ [.1cm]
\multicolumn{3}{l}{\emph{Random Effects: Spatio-temporal field}} \\
  $\quad$  $\theta_1^\tau$  & -0.055\,\,\, (-0.864, 0.649)  &  -0.482\,\,\,  (-1.052, 0.034)\\
  $\quad$ $\theta_2^\tau$  & \textbf{-2.971}\,\,\, (-3.208, -2.701) & \textbf{-2.159}\,\,\,  (-2.444, -1.803) \\
  $\quad$  $\theta_1^\kappa$  & -0.304\,\,\, (-1.386, 0.932) & 0.107\,\,\,  (-0.280, 0.544) \\
 $\quad$  $\theta_2^\kappa$ & -0.120\,\,\, (-0.360, 0.111) & \textbf{-0.113}\,\,\,  (-0.237, -0.031)\\
        $\quad$ Temporal correlation $a$ & \textbf{0.969}\,\,\, (0.945, 0.983) & \textbf{0.886}\,\,\,  (0.765, 0.946)\\ [.1cm]
        \hline
    \end{tabular}
     \caption{Posterior estimates of fixed and random effects for maximum daily precipitation sum normals. Statistically significant effects are highlighted in bold.}
\end{table}


\begin{table}[ht]
    \centering
    \label{tab:posterior_dry_spell}
    \begin{tabular}{lcc}
        \hline
        \textbf{Variable} & \textbf{1961--1990: Mean (95\% CI)} & \textbf{1991--2020: Mean (95\% CI)} \\
        \hline
        \multicolumn{3}{l}{\emph{Fixed Effects}} \\
        \quad Slope & -0.002 (-0.007, 0.002) & 0.003 (-0.001, 0.006) \\
        \quad scale(Aspect) & 0.001 (-0.008, 0.011) & \textbf{0.006} (0.000, 0.012) \\
        \quad scale(Elevation): January & \textbf{-0.058} (-0.078, -0.038) & \textbf{-0.075} (-0.095, -0.054) \\
        \quad scale(Elevation): February & \textbf{-0.080} (-0.100, -0.059) & \textbf{-0.080} (-0.101, -0.059) \\
        \quad scale(Elevation): March & \textbf{-0.086} (-0.106, -0.066) & \textbf{-0.072} (-0.093, -0.052) \\
        \quad scale(Elevation): April & \textbf{-0.080} (-0.101, -0.059) & \textbf{-0.058} (-0.080, -0.037) \\
        \quad scale(Elevation): May & \textbf{-0.045} (-0.067, -0.023) & \textbf{-0.056} (-0.080, -0.033) \\
        \quad scale(Elevation): June & \textbf{-0.038} (-0.062, -0.014) & \textbf{-0.029} (-0.053, -0.005) \\
        \quad scale(Elevation): July & \textbf{-0.055} (-0.077, -0.032) & \textbf{-0.044} (-0.068, -0.019) \\
        \quad scale(Elevation): August & \textbf{-0.047} (-0.069, -0.025) & \textbf{-0.038} (-0.062, -0.015) \\
        \quad scale(Elevation): September & \textbf{-0.050} (-0.070, -0.030) & \textbf{-0.028} (-0.050, -0.007) \\
        \quad scale(Elevation): October & \textbf{-0.028} (-0.047, -0.009) & -0.019 (-0.040, 0.001) \\
        \quad scale(Elevation): November & \textbf{-0.028} (-0.048, -0.008) & \textbf{-0.032} (-0.052, -0.011) \\
        \quad scale(Elevation): December & \textbf{-0.047} (-0.067, -0.027) & \textbf{-0.054} (-0.075, -0.033) \\[.1cm]
             \multicolumn{3}{l}{\emph{Random Effects: Distribution }} \\
\quad Dispersion parameter $n$ & \textbf{12.997} (12.640, 13.353) & \textbf{11.760} (11.415, 12.113)\\[.1cm]
        \multicolumn{3}{l}{\emph{Random Effects: Spatial field}} \\
        \quad $\theta_1^\tau$ & \textbf{2.807} (2 163, 3.390) & \textbf{-0.980} (-1.854, -0.139) \\
        \quad $\theta_2^\tau$ & \textbf{-2.246} (-3.161, -1.1076) & \textbf{1.283} (0.896, 1.691) \\
        \quad $\theta_1^\kappa$ & 1.165 (-0.535, 3.439) & \textbf{1.057} (0.471, 1.666) \\
        \quad $\theta_2^\kappa$ & -0.502 (-5.480, 1.474) & \textbf{-0.743} (-1.103, -0.403) \\[.1cm]
        \multicolumn{3}{l}{\emph{Random Effects: Spatio-temporal field}} \\
        \quad $\theta_1^\tau$ & \textbf{-0.737} (-1.365, -0.181) & -0.084 (-0.486, 0.341) \\
        \quad $\theta_2^\tau$ & \textbf{0.927} (0.563, 1.251) & \textbf{1.316} (1.080, 1.549)  \\
        \quad  $\theta_1^\kappa$ & -0.113 (-0.714, 0.562) & -0.473 (-1.008, 0.025)\\
        \quad $\theta_2^\kappa$ & \textbf{-0.091} (-0.174, -0.013) & -0.083 (-0.168, 0.003) \\
        \quad Temporal coefficient $a$ & \textbf{0.947} (0.900, 0.979) & \textbf{0.714} (0.582, 0.823)\\  [.1cm]
        \hline
    \end{tabular}
     \caption{Posterior estimates of fixed and random effects for maximum length of a dry spell normals. Statistically significant effects are highlighted in bold.}
    \end{table}

The Elevation:Month interaction term reveals complex and noteworthy seasonal shifts. For mean precipitation sums, the most striking finding is the increased positive influence of elevation during winter months January and February in the more recent 1991-2020 period. 
A one-standard-deviation increase in elevation is associated with a $13.6\%$ increase in the expected normal mean precipitation in January and $16.1\%$ in February. This indicates that higher altitudes in Austria are experiencing a greater increase in winter relative to lower elevations, amplifying their role as precipitation receivers in colder months. Conversely, the statistically non-significant effect of elevation on mean precipitation sums during summer and autumn months (June to October) in the later climate period is noteworthy. This shows that during these months, the enhancing effect of mountains on precipitation is less pronounced compared to 1961-1990. 
For maximum precipitation, elevation maintains a strong positive effect across all months. However, its magnitude generally decreases during the late spring and summer months May-August in the later period, while significantly increasing in September, October and December. This points to a seasonal redistribution of the elevation-dependent extreme rainfall. For maximum dry spell length, elevation generally has a negative effect (shorter dry spells at higher elevations), but this 'protective' effect weakens across many months in the later period (e.g., from $-8.6\%$ to $-7.2\%$ in March, from  $-8\%$ to $-5.8\%$ in April and from $-5\%$ to $-2.8\%$ in September), suggesting that high-altitude areas are experiencing relatively longer dry spells than before.

Across all precipitation characteristics, the slope consistently exhibits no statistically significant effect in either time period.
This suggests that terrain steepness itself does not play a quantifiable linear role.
Conversely, the aspect of the terrain exhibits a statistically significant influence. For monthly mean precipitation, the effect of aspect more than doubles in magnitude in the 1991-2020 period (from $-1.2\%$ to $-2.6\%$), indicating a significantly greater reducing impact.   
Similarly, for monthly maximum precipitation, the negative aspect effect intensifies (from $-15.6\%$ to $-26.9\%$). This collective strengthening implies that slope direction is becoming a more critical determinant for both mean and maximum precipitation. For dry spell length, aspect, while non-significant in the earlier period, becomes statistically significant and positive ($0.6\%$) in the later period, indicating that certain aspects are now associated with longer dry spells.\footnote{Recall that the aspect and elevation were standardised, such that the coefficients can directly be interpreted as effect sizes.}

The distribution parameter estimates indicate a slight but statistically significant increase in overall variability for the later period: the precision parameter $\phi$ for mean precipitation (from $3.014$ to $2.852$) and the dispersion parameter $n$ for dry spell length (from $12.997$ to $11.760$) decline, while the spread parameter $s_\beta$ for maximum daily precipitation increases (from $4.675$ to $4.940$).

Regarding non-stationary effects and temporal correlation, the spatial field parameter $\theta_2^\kappa$  indicates a overall negative influence of elevation on the spatial correlation range for any precipitation characteristics. Only for dry spells the spatial field is significantly non-stationary in the later period.  
The temporal correlation coefficient $a$ reveals distinct shifts: for mean precipitation, it increases (from $83.1\%$ to $87.3\%$). However, for maximum daily precipitation, it decreases (from $96.9\%$ to $88.6\%$), and most notably for dry spell length, it significantly decreases (from $94.7\%$ to $71.4\%$). 
This suggests that while mean precipitation is becoming more persistent, extremes and dry spells exhibit less temporal consistency, implying more dynamic  variations in these characteristics.

The computation times were 34 min and 53 min to run the prediction of mean precipitation normals,  41 min and 43 min for the normals of maximum length of a dry spell and 67 min and  88 min for the maximum daily precipitation sums in the early and late time period, respectively.

\subsection{Cross-validation metrics}\label{sec:cv}

\begin{figure}[h!]
    \centering
    \includegraphics[scale=0.35]{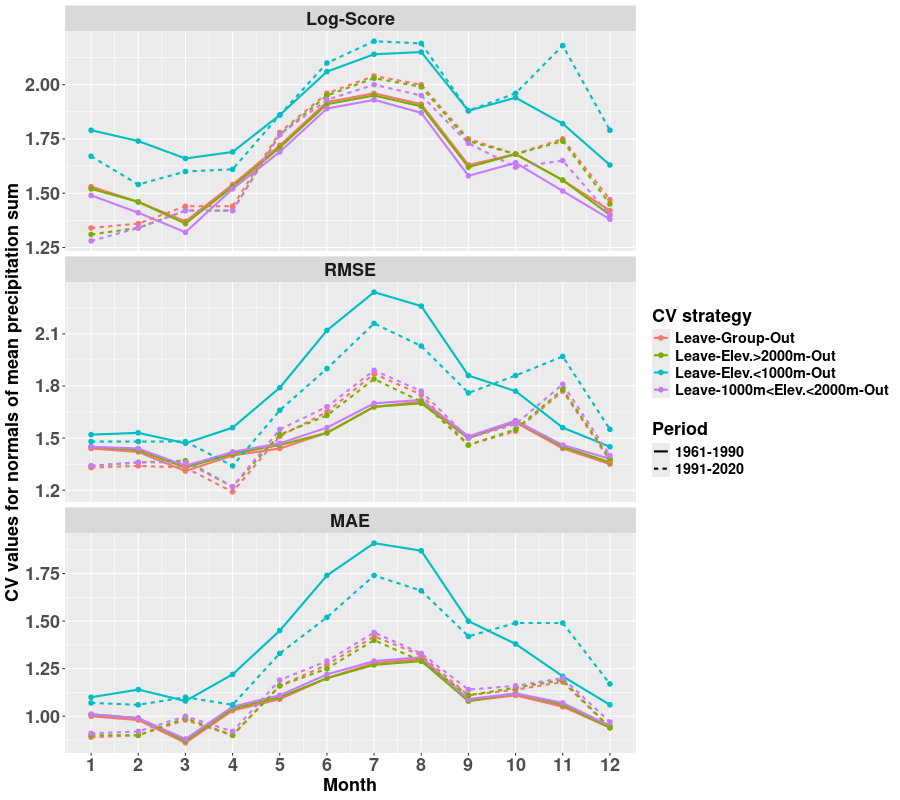}
    \caption{Comparison of CV values for Leave-Group-Out, and Leave-Elevation range-Out for monthly normals of mean precipitation sums.}
    \label{fig:cv_mean}
\end{figure}

To comprehensively assess the predictive performance of the precipitation models in the two climatological reference periods, we employ various cross-validation (CV) strategies. The validation methods include: Leave-Group-Out (LGO) CV for spatially structured models (\cite{Adin2024}), and a custom Leave-Elevation range-Out approach developed specifically for this modelling setup. 
The performances are quantified using Root Mean Square Error (RMSE), Mean Absolute Error (MAE), and Log-Score. Lower values for both the RMSE and MAE are preferred, as they indicate a closer match between the predicted and observed data. In contrast, for the Log-Score, a value closer to 0 is desired.\\
For mean precipitation sums, the Leave-Elevation$<$1000m-Out CV  strategy yields the highest errors across all three metrics, seen in Figure \ref{fig:cv_mean}. 
Obviously, the model struggles to predict the mean precipitation sums in low-elevation areas, without available data from these regions.
 For both periods, the errors peak in July, but in the 1991-2020 period, a second rise in errors is also observed toward the end of the year.
Also notable is that errors from other Leave-Elevation range-Out strategies are more pronounced during the summer months in the later climate normal period (1991–2020) than in the earlier one (1961-1990).

\begin{figure}[h!]
    \centering
    \includegraphics[scale=0.35]{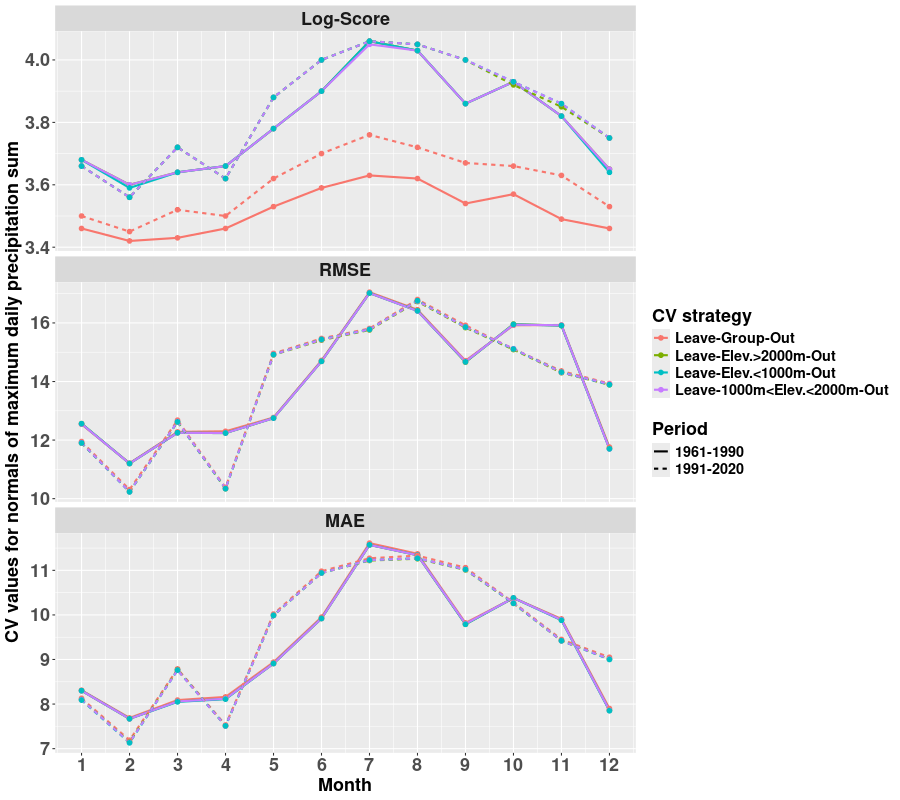}
    \caption{Comparison of CV values for Leave-Group-Out, and Leave-Elevation range-Out for monthly normals of maximum daily precipitation sums.}
    \label{fig:cv_max}
\end{figure}

Figure \ref{fig:cv_max} presents the CV values for the normals of maximum daily precipitation sums. While all strategies yield the same error plot for RMSE and MAE, the Leave-Group-Out approach shows the lowest Log-Score among all CV strategies.
 A common temporal pattern is observed across all metrics: errors are lowest at the beginning of the year, increase from April onward to a peak in either July or August, and then decrease toward the end of the year. Compared to the other precipitation characteristics the extreme value modelling of monthly maximum daily precipitation values shows the highest CV errors.

\begin{figure}[h!]		
 \centering
 \includegraphics[scale=0.4]{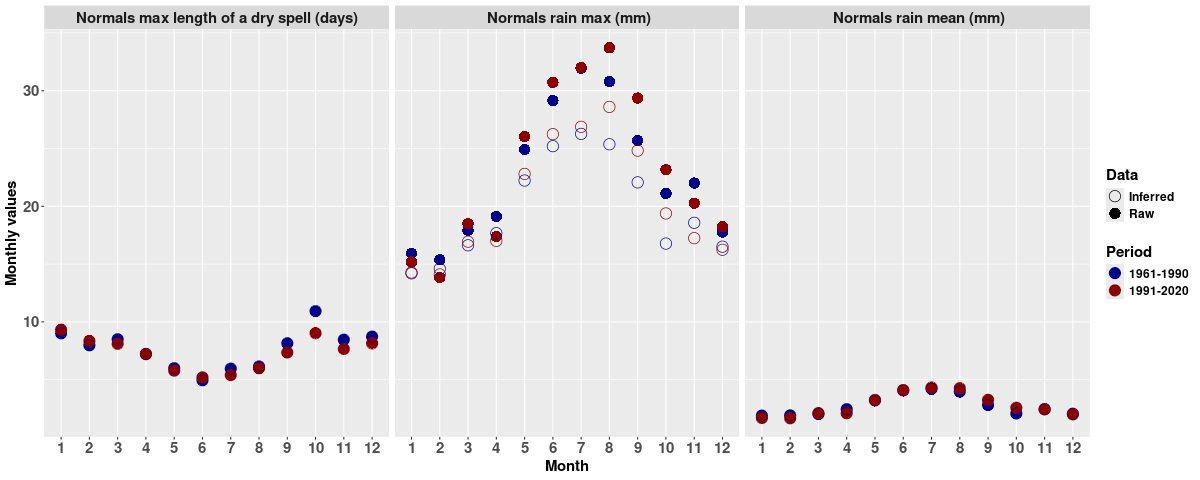}
		\caption{Comparison of observed and inferred values of monthly normals for three precipitation characteristics across the 1961-1990 and 1991-2020 climate normal periods.}
		\label{fig:normals}
\end{figure}

 This is further reflected in Figure \ref{fig:normals}, which compares inferred and observed normals and reveals a tendency for the extreme value model to underestimate monthly values, particularly from May onward.
Given that the 20-year return values are computed directly from the inferred monthly maximum daily precipitation sums, these discrepancies are particularly crucial. The model's tendency to underestimate these maxima, therefore directly impacts the accuracy of the estimated amplitude of the return values,  underscoring the inherent difficulty in modelling extreme values accurately.

\begin{figure}[h!]
    \centering
    \includegraphics[scale=0.35]{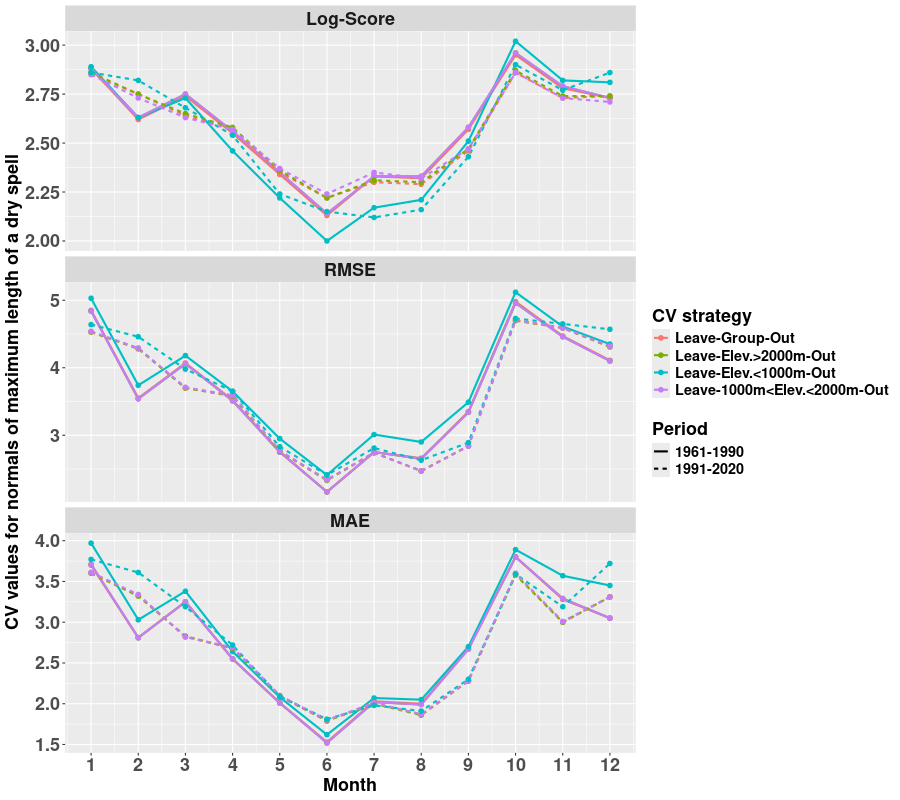}
    \caption{Comparison of CV values for Leave-Group-Out, and Leave-Elevation range-Out for monthly normals of maximum dry spell length.}
    \label{fig:cv_dry}
\end{figure}

While the errors for monthly mean precipitation sum normals are highest in summer, the CV values for the normals of monthly maximum length of a dry spell show a U-shaped temporal pattern across all three metrics, see Figure \ref{fig:cv_dry}. However, the error pattern for the later period (1991-2020) is notably smoother at the beginning of the year compared to the early time period.

\section{Discussion}

A critical aspect employed in this study, is the specification of prior distributions for model parameters and hyperparameters (as detailed in Subsection \ref{sec:priors}). Future research could systematically explore the robustness of our findings to alternative prior specifications. Such detailed research into prior sensitivity would not only serve to confirm the reliability of our current conclusions but could also offer deeper insights into the complex interplay between data, model structure, and prior assumptions in the spatio-temporal modelling of precipitation characteristics.

A recognised challenge, particularly in the modelling of extreme precipitation events such as monthly maximum precipitation sums, lies in the complexity of fitting appropriate distributions. Our use of the bGEV distribution, while theoretically robust for capturing extreme value behaviour, presents computational difficulties during fitting. 
Future work could explore alternative, perhaps computationally more stable, extreme value distributions to ensure greater reliability of these particular estimates.

Another potential area for model enhancement involves the assumption of separability in the spatio-temporal covariance structure. Our current model assumes that the spatial and temporal dependencies of precipitation can be modelled independently. While this assumption offers a significant advantage by substantially reducing computational complexity, it may not fully capture the reality of precipitation patterns in a complex mountainous region like Austria. 

\section{Conclusion}
This paper presents a statistical model capturing local and region-specific orographic influences on precipitation characteristics in Austria for the climatic reference periods 1961-1990 and 1991-2020. The model enables comparison of monthly precipitation normals across Austria’s administrative borders on a 2$\times$2 km grid.


We split the approach in  three key tasks. Firstly, we identify  and process potential covariates influencing precipitation before specifying distributional assumptions for the three precipitation characteristics: monthly mean and maximum daily precipitation sums as well as the monthly maximum length of a dry spell in Austria. These precipitation indices are modelled using the gamma, bGEV and negative binomial distribution, respectively.
Secondly, we describe the data composition through a generalised linear mixed model, incorporating slope, aspect, an interaction between elevation and month, a spatial non-stationary GRF and a spatio-temporal GRF over two standard reference periods. 
Lastly, we use Bayesian inference and \texttt{R-INLA} in order to explore how precipitation patterns evolve over the last 60 years.
The posterior maps of the precipitation changes illustrate relative differences over time, showing that mean precipitation normals generally declined early in the year  but increased notably in March in northern Austria and across the country in September and October ($\sim +25-50\%$).
The maximum length of a dry spell extended significantly in January, February, and June, particularly in southern regions (up to $+30\%$).
Maximum daily precipitation amounts rose markedly in August through October nationwide (up to $+30\%$), while 20-year return values increased more moderately, by up to $10\%$.
Our parameter estimation reveals that both the aspect as well as the interaction between elevation and month reveal statistically significant insights in the change of the precipitation characteristics.
We evaluate our estimates using several CV strategies, including a custom-designed Leave-Elevation range-Out approach. Among all precipitation characteristics, the CV of monthly maximum daily precipitation sums modelled with a bGEV extreme value distribution yields the highest error, which highlights the inherent challenge of accurately modelling such extreme events.

While developed for Austria, the modelling approach presented in this study is broadly applicable beyond the national context. Relying on publicly available geographic covariates, the methodology can be readily transferred to other countries, supported by the global availability of topographical data through platforms such as GADM.

\section{Data availability statement}
The weather data provided by GeoSphere  can be downloaded from \url{https://dataset.api.hub.geosphere.at/app/frontend/station/historical/klima-v2-1d}. The elevation map is taken from the Global Administrative Areas and can be found at \url{https://gadm.org/download_country.html}.    The pre-processed data and code are available at \url{https://github.com/CorinnaPerchtold/Climate_Change}.

\section{Acknowledgment}
Special thanks to Finn Lindgren and Havard Rue for their insightful discussions in the 'R-INLA' Google group, and to Johan Lindström for his guidance in setting up the code and for hosting me at Lund University through the Erasmus+ program. We also thank Manuel Kauers from the Institute of Algebra at JKU for letting us perform the computations reported in this paper on the computers of his group. Also thanks to Helga Wagner for giving me great advice on my paper. We also extend our gratitude to the Editors and reviewers for their valuable feedback, which significantly enhanced the quality of this article.

\section{Supplementary material}\label{sec:Appendix}

\begin{figure}[h!]
       \centering
    \includegraphics[scale=0.3, trim=10mm 0 0 0,clip]{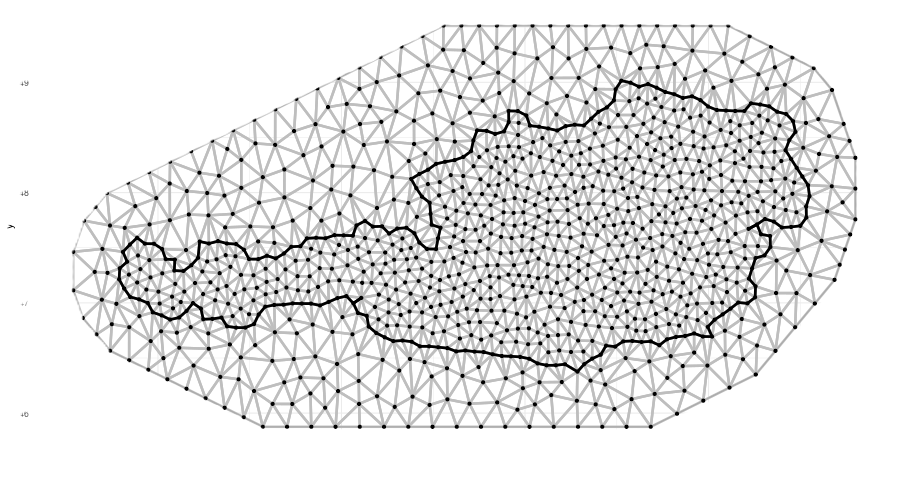}
       \caption{Triangulation of Austria with $1091$ mesh vertices.}
       \label{fig:mesh}
   \end{figure}

\begin{figure}[h]
    \centering
\includegraphics[scale=0.3]{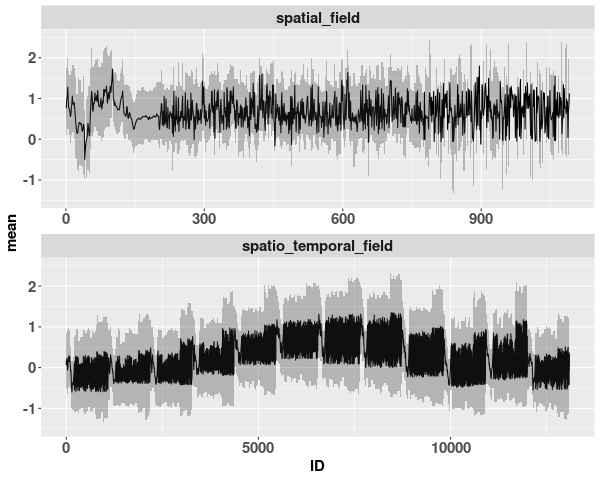}
\includegraphics[scale=0.3]{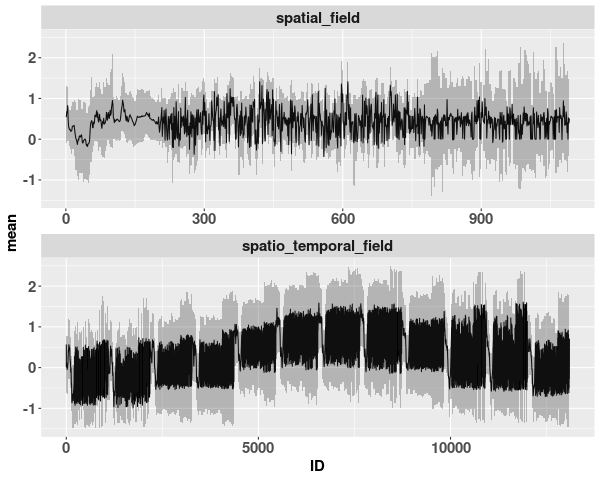}
    \caption{ Marginal posterior mean and $95\%$ credible intervals of the
spatial and spatio-temporal random effects of monthly mean precipitation sums. Left: 1961-1990. Right: 1991-2020.}
    \label{fig:GMRF_mean}
\end{figure}

\begin{figure}[h!]
    \centering
    \includegraphics[scale=0.3]{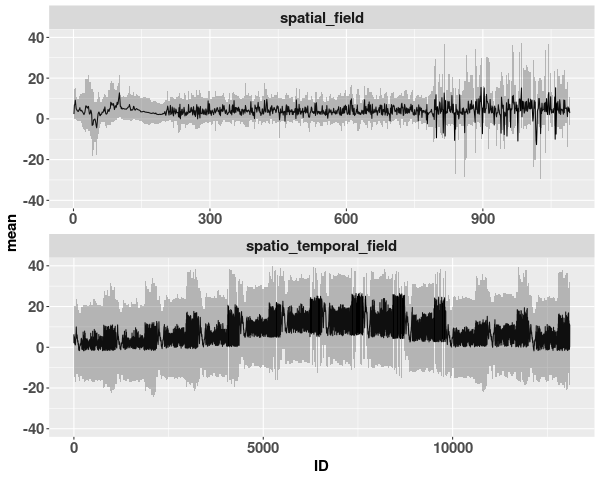}
        \includegraphics[scale=0.3]{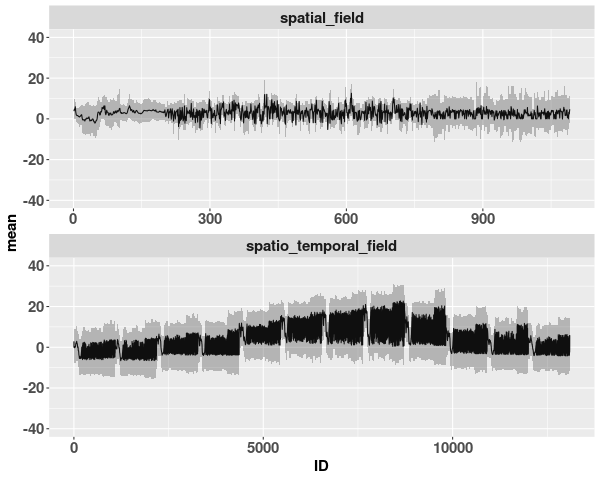}
    \caption{ Marginal posterior mean and $95\%$ credible intervals of the
spatial and spatio-temporal random effects of monthly maximum  daily precipitation sums. Left: 1961-1990. Right: 1991-2020.}
    \label{fig:GMRF_max}
\end{figure}

\begin{figure}[h!]
    \centering
    \includegraphics[scale=0.3]{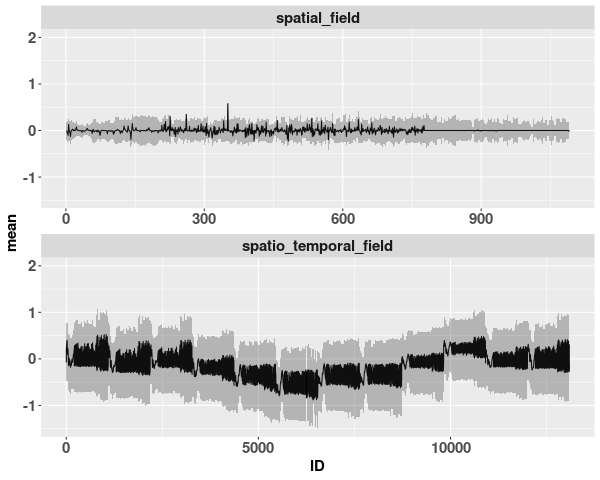}
    \includegraphics[scale=0.3]{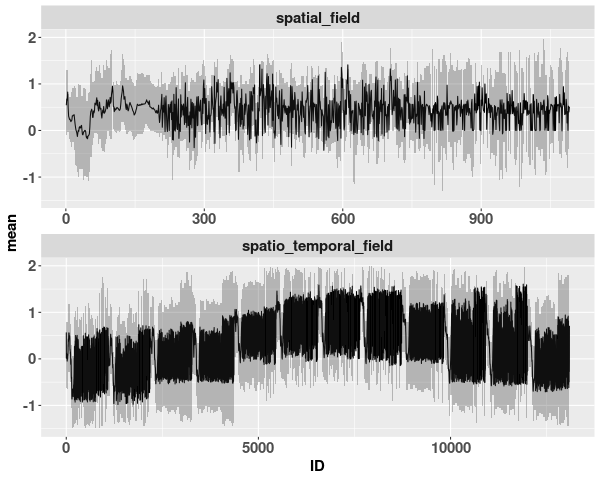}
    \caption{ Marginal posterior mean and $95\%$ credible intervals of the
spatial and spatio-temporal random effects of monthly maximum length of a dry spell. Left: 1961-1990. Right: 1991-2020.}
    \label{fig:GMRF_min}
\end{figure}

We present the marginal posterior mean and its $95\%$ credible interval of the spatial and spatio-temporal random fields, corresponding to each time period and precipitation scenario. 
Specifically, the upper panels in Figures \ref{fig:GMRF_mean}, \ref{fig:GMRF_max} and    \ref{fig:GMRF_min} display the posterior mean and 95\% credible interval (CI) of the corresponding spatial random effect $\delta(s_j)$, which is modelled as a GMRF triangulated over $j=1, \ldots, 1091$ mesh points for each of the two climate normal periods.
 Generally, the variability of the posterior mean increases for mesh points outside of Austria’s borders, see the outer mesh in Figure \ref{fig:mesh}.
 
Similarly, the lower panels, presented for each scenario, illustrate the posterior mean and its 95\% CI of the spatio-temporal random effect $z(s_j,t_k)$, modelled also as a GMRF, over $j=1, \ldots, 1091$ mesh points and $k=1,\ldots, 12$ months. These prediction points are denoted by ID in the figures.  The spatio-temporal random effect is temporally disaggregated, with each block illustrating the spatial random effect $w(s_j)$ for a distinct month.

The mean and 95\% CI of the spatial field of mean precipitation sum normals appear relatively consistent between the 1961-1990 and 1991-2020 periods. This suggests that the unexplained local influences on mean precipitation have not drastically changed their characteristics over the 60 years. Similarly,  the overall amplitude and seasonal pattern of the spatio-temporal field appears rather unchanged  between the two periods. However, a slight increase in the width of the 95\% CI in the spatio-temporal field can be noted, see  Figure \ref{fig:GMRF_mean}. 

For the normals of monthly maximum daily precipitation sums, the posterior mean and its 95\% CI of the spatial field appear to be more variable in the early time period than in the late one. Also the seasonal pattern in the lower panel is more pronounced in the early time period, as seen in Figure \ref{fig:GMRF_max}. The smoothness in the late spatial estimates is also evident within the spatio-temporal field of the same period. 

The estimates of the random effects for the normals of monthly maximum length of a dry spell, show diverse pictures from one climate normal period to the other. While the spatial variability appears to be around zero in the early period, it is higher in the later period.
A reversing seasonal trend can be identified in the spatio-temporal field from 1961-1990 to 1991-2020, see Figure \ref{fig:GMRF_min}.

\newpage

\bibliographystyle{plain}
		\bibliography{lib_tex.bib}

\end{document}